\documentclass[journal]{IEEEtran}
\usepackage{balance}
\usepackage[T1]{fontenc}                       
\usepackage{graphicx}
\usepackage[justification=centering]{caption}  
\usepackage{booktabs}                          

\begin{document}

\title{Can We Trust AI in 6G? Verifiable and Auditable AI-Driven Trustworthy Wireless Networks}

\author{Genze Jiang, Yizhou Huang, Kezhi Wang, \textit{Senior Member, IEEE}

\thanks{Corresponding author: Kezhi Wang}
\thanks{Genze Jiang, Yizhou Huang and Kezhi Wang are with the Department of Computer Science, Brunel University London, UK (e-mail: Genze.Jiang@brunel.ac.uk,  Yizhou.Huang2@brunel.ac.uk, Kezhi.Wang@brunel.ac.uk).}}

\markboth{Submitted for Review}%
{Shell \MakeLowercase{\textit{et al.}}: Bare Demo of IEEEtran.cls for IEEE Journals}
%

\maketitle

\begin{abstract}
Mobile network operators are increasingly exploring the use of artificial intelligence (AI) to automate network tasks such as cell selection and mobility management. Current assurance methods for these AI functions are forms of output testing. They check what a function decides, but they cannot show why it decided that way, and two functions can pass the same tests for different reasons. In this paper, we propose mechanistic auditing, which inspects the internal features of an AI function and checks them against machine-verifiable 3GPP specifications. We set out a three-step auditing principle that locates protocol-relevant features, verifies whether decisions depend on them, and audits each decision against them. We then present an audit-native network architecture. In this architecture, a dedicated verification agent certifies each AI function before deployment and audits its decisions during operation. In a case study on a language model for LTE signalling, we find internal features that carry the 3GPP handover concepts. Removing them lowers the model's accuracy on handover tasks from 96\% to 0\%, while removing the same number of random features changes its accuracy by less than two points. An audit of a handover decision can therefore check whether these features are active when the model makes that decision. We also discuss the open challenges that remain before mechanistic auditing can enter telecommunications practice and standardisation.
\end{abstract}

\begin{IEEEkeywords}
Mechanistic auditing, trustworthy AI-driven wireless networks, large language model, AI agents, 6G, protocol compliance, network reliability.
\end{IEEEkeywords}

\section{Introduction}
\label{sec:intro}

\IEEEPARstart{M}{obile} network operators are exploring artificial intelligence (AI) to automate tasks such as mobility management, network slicing, and resource allocation~\cite{chen2024big}. 3GPP Releases~18 and~19 reflect growing efforts to support AI and machine learning (ML) in wireless systems. Alongside these developments, recent studies examine large language models (LLMs) for telecommunications~\cite{zhou2024large}, including agents that interpret network state and coordinate network functions~\cite{xu2024large}. Telecommunications benchmarks reveal limitations in these models' domain knowledge~\cite{maatouk2025teleqna}, highlighting the need for reliable evaluation as their use in network applications expands.

AI network functions can be evaluated through conformance testing, shadow-mode deployment, canary rollout, and constraint mechanisms that reject non-compliant outputs. These methods primarily assess observable behaviour and network outcomes without directly establishing whether a model's decisions depend on the relevant protocol conditions. Models with similar test accuracy may rely on different cues, including protocol conditions or correlations specific to the training data, and may therefore respond differently when operating conditions change.

We propose \emph{mechanistic auditing} to complement output testing of language-model-based network functions with evidence from their internal computation. The audited \emph{AI function} is the model-based component that supports or makes a network decision, whether deployed independently or within an agent. The audit identifies internal features that are associated with relevant protocol concepts, where a feature is a component of the model's activation representation. Controlled interventions then assess the contribution of these features to task performance. Before deployment, feature alignment and intervention tests establish a baseline linking selected internal features to protocol concepts. During operation, the audit checks for each decision whether the features that carry the governing concepts are active and whether the decision follows them. A decision that fails this check raises an alert or is blocked.

The approach combines \emph{mechanistic interpretability} with specification-based network assurance. Mechanistic interpretability studies the internal computations that produce a model's outputs~\cite{bereska2024mechanistic}. Sparse autoencoders decompose model activations into features whose associations with protocol concepts can be evaluated~\cite{huben2024sparse}. Recent studies have extended this approach to large models~\cite{templeton2026scaling} and developed benchmarks for evaluating the extracted features~\cite{karvonen2025saebench}. The need for such a decomposition restricts mechanistic auditing to language-model-based functions, because the regression, tree-based and reinforcement-learning functions common in radio access networks do not admit it. Selected conditions from 3GPP specifications are translated into executable predicates that assign true or false labels to signalling scenarios. These labels provide an external reference for evaluating associations between internal features and protocol concepts.

This article develops a three-step auditing principle that links specification-derived protocol conditions to internal feature analysis, and proposes an architecture that incorporates the resulting evidence into network operation. The case study in Section~\ref{sec:casestudy} applies the principle to Qwen3-0.6B on questions derived from recorded LTE signalling, with selected comparisons across eight language models, 5,000 signalling tasks and 15 concepts, 12 of which are defined over the measurements and events of TS~36.133 and TS~36.331.


\begin{figure*}[!t]
\centering
\includegraphics[width=\textwidth]{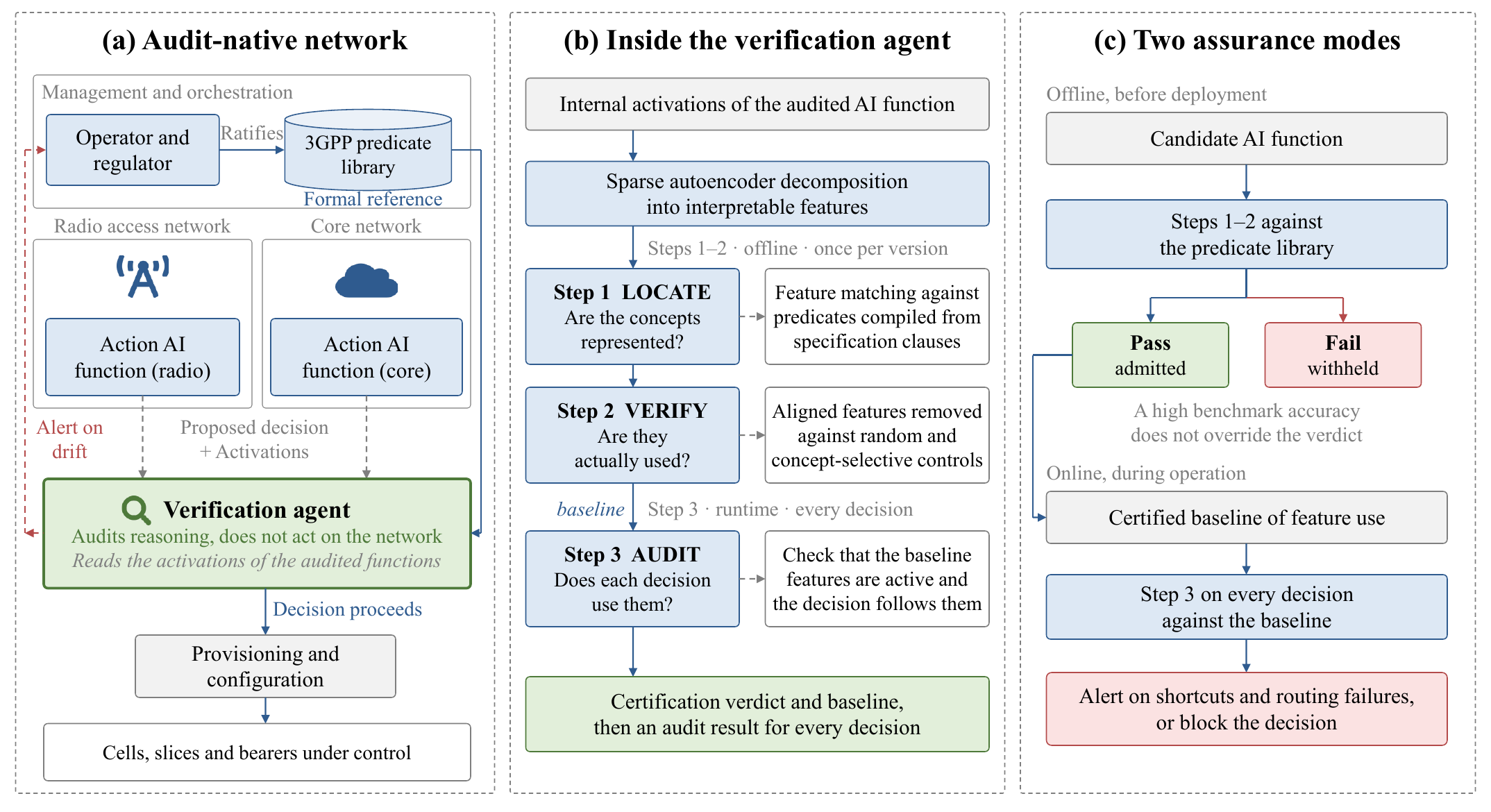}
\caption{The proposed audit-native network. (a) A verification agent audits the AI functions before their decisions reach provisioning. (b) The three-step audit runs inside that agent. (c) The two assurance modes, pre-deployment certification and runtime auditing.}
\label{fig:overview}
\end{figure*}

\section{Why Networks Demand Auditing}
\label{sec:why}

Current assurance methods for AI functions are forms of output testing. They check what a function decides, but they cannot show why it decided that way. This holds even for a rule-based checker that compares every decision with the 3GPP specification, because the checker sees only the decision and not the reasoning behind it. Two functions can therefore pass the same tests for different reasons. One follows the protocol conditions in the specification. The other relies on a pattern that happened to hold in its training data, and it may fail when conditions change. Output testing cannot tell them apart. To do so, we should look inside the function and check whether it represents the relevant protocol concepts and whether its decisions depend on them. Mechanistic auditing performs this check.

The limitation of output testing is especially serious in wireless networks, for two reasons. First, the networks that these AI functions control also carry safety-critical services such as emergency response and industrial control. A decision that is correct for the wrong reason may fail when conditions change, and the failure then affects these services directly. Second, networks are moving towards several AI agents that work together, with the output of one agent becoming the input of the next~\cite{maatouk2024large}. If one agent reasons incorrectly, its error passes to every agent after it, while the final output may still look plausible. A fault in one place can then disrupt a whole service. Output testing finds such a fault only after it has caused harm, and auditing the internal reasoning of each function can find it earlier. Fig.~\ref{fig:overview} previews our proposal.

\section{Three-Step Auditing Principle}
\label{sec:principle}

We now set out a general principle for mechanistic auditing. The principle has three steps, summarised in Fig.~\ref{fig:overview}(b). The first two steps examine the function as a whole. They ask whether the protocol concepts relevant to its decisions are represented inside it and whether its decisions depend on them. They run offline, once for each version of the function, and they produce a baseline that records which internal features carry each protocol concept. The third step asks the same two questions of each individual decision, using the baseline to know which features to read. The first two steps therefore support pre-deployment certification, and the third supports runtime auditing.

\subsection{Step 1: Locate Protocol-Relevant Internal Features}

The first question is whether the protocol concepts relevant to a decision are represented inside the AI function at all. Answering it requires a way to translate an AI function's internal activations into human-interpretable units. Sparse autoencoders have emerged as a leading tool for this purpose, re-expressing the dense activations of a network as a sparse set of features that tend to be more interpretable than individual neurons~\cite{huben2024sparse}. Sparse probing of internal activations has likewise been used to test whether specific concepts are linearly recoverable from a model's representations~\cite{gurnee2023finding}. Auditing then asks which of these features carry a protocol concept, for example a signal-quality threshold or the condition that triggers a handover.

\begin{figure*}[!t]
\centering
\includegraphics[width=\textwidth]{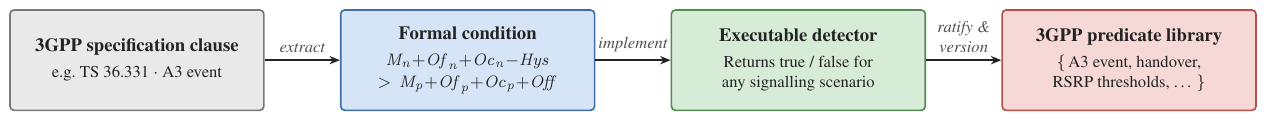}
\caption{Constructing the predicate library. A 3GPP clause with a numeric decision condition is compiled into an executable detector that returns true or false for any signalling scenario. The audit checks features against the ratified, versioned collection of detectors.}
\label{fig:predlib}
\end{figure*}

To find these features, we turn each protocol concept into a detector, as Fig.~\ref{fig:predlib} shows. A detector starts from a clause of a 3GPP specification, for example the clause of TS~36.331 that defines the A3 event for handover. The clause states a condition on measured values, shown as the boxed formula in the figure. The formula means that the network considers a handover only when the neighbour cell's measured signal, after its frequency and cell offsets, exceeds the serving cell's by more than a hysteresis margin plus an event offset. That required margin keeps a brief fluctuation from triggering an unnecessary handover. We write the condition as a short program. Given the measurements in a scenario, the program returns true when the condition holds and false when it does not. The collection of these detectors forms the predicate library.

The detector tells us, for each test scenario, whether the A3 condition holds. We use these labels to find the features that carry the A3 concept. We give the AI function many signalling scenarios, and the A3 condition holds in some of them and fails in the others. For each scenario, we record how strongly each feature activates. A feature that carries the A3 concept is strongly active when the condition holds and weak when it fails. A statistical test checks that this difference is too large to arise by chance among the many features and concepts compared. Step~1 thus produces, for each protocol concept, a list of the features that carry it. We call these the aligned features of the concept.

\subsection{Step 2: Verify the Causal Role of Features}

A feature that carries a protocol concept may still play no part in the decision. The second question is therefore whether the decision depends on these features. We test this by removing the features from the internal computation. If the decision depends on them, removing them should change it. Removing unrelated features should leave it unchanged. Such interventions are well established in interpretability research. Editing a localised internal association can change a model's factual output in a controlled way~\cite{meng2022locating}, tracing the features that take part in a behaviour recovers the causal subgraph behind it~\cite{marks2025sparse}, and steering along identified directions can predictably shift model behaviour at inference~\cite{zou2023representation}. This step separates an AI function that relies on a protocol concept from one that reaches the right answer through an unrelated shortcut.

Removing many features at once harms almost any model. A drop in accuracy after removing the aligned features therefore proves little on its own, and we compare it against two controls. The first control removes the same number of features, chosen at random from the same layer. The aligned features matter only if removing them harms the decisions much more than removing random features does. The second control checks that the harm is specific to the concept. It removes only the aligned features of one concept, such as A3, and measures the effect on the handover decisions and on all other decisions. If these features are specific to the A3 concept, the handover decisions suffer much more than the others. When the aligned features pass both controls, the function's decisions depend on them.

Steps 1 and 2 together produce the certification verdict for the function. The function passes when three conditions hold for every protocol concept relevant to its decisions. First, the concept has at least one aligned feature. Second, removing these features changes the decisions that the concept governs significantly more than the random control does. Third, the effect is specific to the concept. The aligned features of a certified function form the baseline for the third step. The baseline stays fixed until the function is updated, and Steps 1 and 2 are then repeated.

\subsection{Step 3: Audit Individual Decisions}

\begin{figure}[!b]
\centering
\includegraphics[width=\columnwidth]{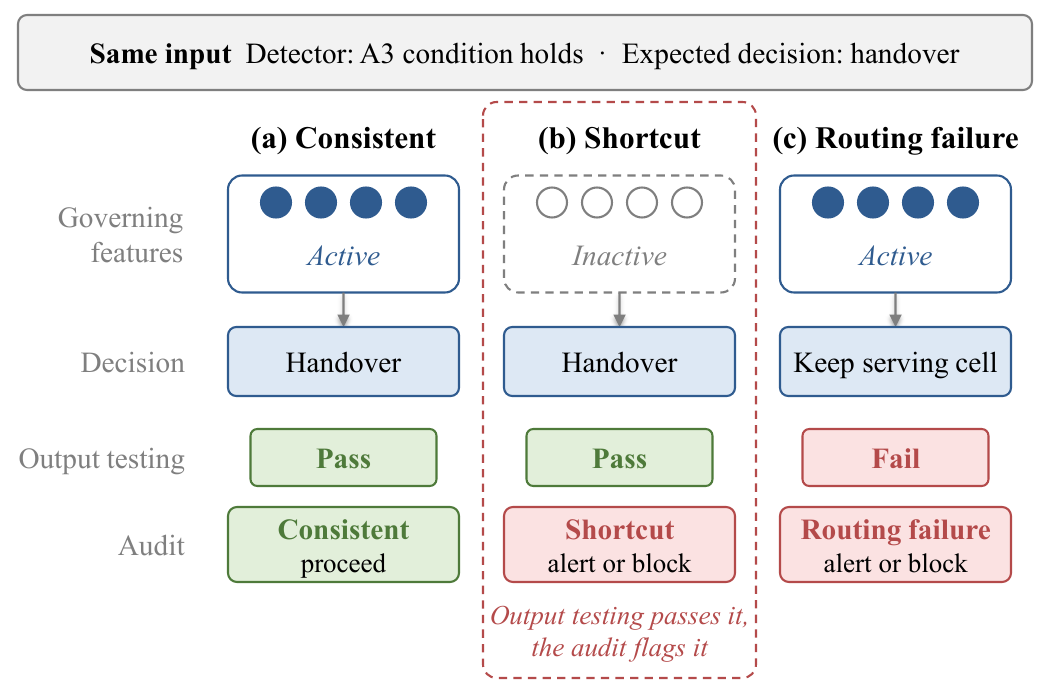}
\caption{Example of the three cases of Step~3 for a handover decision.}
\label{fig:cases}
\end{figure}

The third step runs on every decision that the function makes, using the baseline that the first two steps produced. For each decision, the detectors check the input and report which protocol conditions hold in that scenario. We call the aligned features of these conditions the governing features of the decision. The audit reads their activations and checks two things. The first is whether the governing features are active. The second is whether the decision follows them. For example, when the A3 condition holds, the A3 features should be active and the function should decide to hand over.

Each decision therefore falls into one of three cases. In the first, the governing features are active and the decision follows them, which is the pattern that certification established. In the second, the governing features are inactive although their conditions hold. The function reached its decision without using the protocol condition. In the third, the governing features are active but the decision contradicts them. The function detected the condition but did not act on it. We refer to the second case as a shortcut and to the third as a routing failure. Output testing cannot see either case when the decision happens to be correct. Fig.~\ref{fig:cases} shows an example when the A3 condition holds. In the shortcut case of Fig.~\ref{fig:cases}(b), the function decides to hand over while its A3 features are inactive. Output testing passes this decision, but the audit flags it. Section~\ref{sec:architecture} uses these cases to decide when to raise an alert and when to block a decision.

\section{Audit-Native Network Architecture}
\label{sec:architecture}

We propose to make auditing a component of the network itself. We call this design an \emph{audit-native} architecture. It contains two kinds of agent. An \emph{action agent} is an AI function that perceives network state and acts on it. A dedicated \emph{verification agent} audits the reasoning of the action agents and does not act on the network itself.

\subsection{Separating Action from Auditing}

The verification agent certifies each action agent with Steps 1 and 2 and audits each of its decisions with Step~3. Depending on the mode of the audit, a decision that fails the check either counts towards an alert or is held back from provisioning. Separating the two roles follows the engineering practice of independent verification. Each action agent is checked by a separate component. Fig.~\ref{fig:overview}(a) illustrates the arrangement.

\begin{figure*}[!t]
\centering
\includegraphics[width=\textwidth]{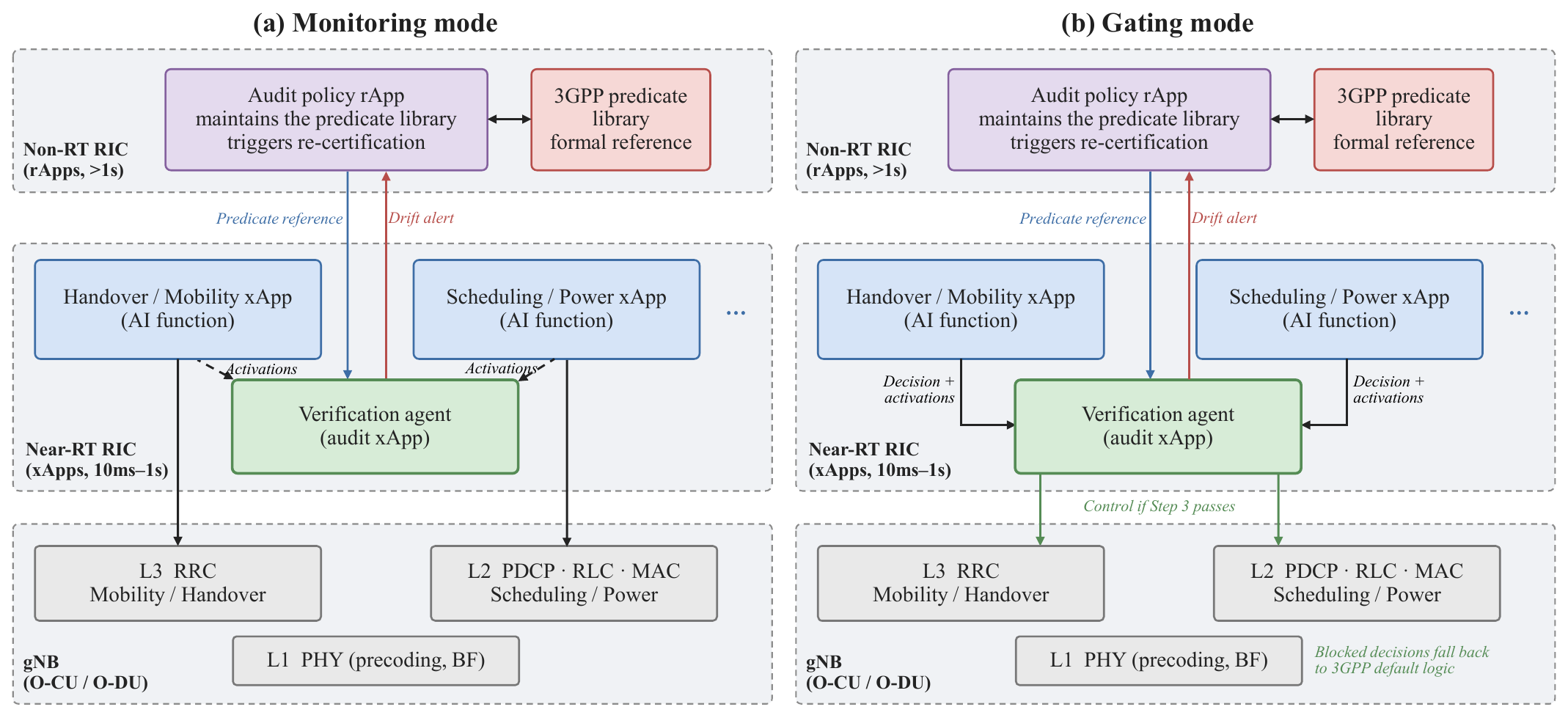}
\caption{The audit-native architecture in an O-RAN base station, shown in its two runtime modes, (a) out-of-band monitoring and (b) in-path gating.}
\label{fig:oran}
\end{figure*}

\subsection{Pre-Deployment Certification and Runtime Auditing}

\begin{table*}[!t]
\renewcommand{\arraystretch}{1.3}
\centering
\caption{The two assurance modes of the audit-native architecture.}
\label{tab:modes}
\begin{tabular}{@{}p{3.0cm}p{7.0cm}p{7.0cm}@{}}
\toprule
\textbf{Aspect} & \textbf{Offline pre-deployment certification} & \textbf{Online runtime auditing} \\
\midrule
When  & Offline, before deployment & Continuously, during operation \\
Steps & Steps 1 and 2, once per version of the function & Step 3, on every decision \\
Questions & Does the function represent the relevant protocol concepts, and do its decisions depend on them? & Are the governing features active, and does the decision follow them? \\
Output & Certification report and baseline & Alert on a rising rate of shortcuts and routing failures, or a blocked decision \\
Role & Keeps functions that reach correct answers through shortcuts out of the network & Detects drift during operation \\
\bottomrule
\end{tabular}
\end{table*}

The audit-native architecture supports two complementary modes of assurance, summarised in Table~\ref{tab:modes} and sketched in Fig.~\ref{fig:overview}(c). The first is \emph{pre-deployment certification}. An action agent is audited offline against the formal predicate library before it is allowed into the network, producing a structured report that records which protocol concepts are internally represented, which of those are causally used, and where gaps remain, together with the baseline used by runtime auditing. The verification agent flags an agent as a shortcut learner when its correct outputs are not backed by causally grounded protocol features, and withholds it regardless of benchmark accuracy. A regulator could in principle accept the certification report as mechanistic evidence, unlike a bare accuracy figure.

The second is \emph{runtime auditing}. Once an action agent is certified, the verification agent holds its baseline and applies Step~3 to each of its decisions. Fig.~\ref{fig:oran} shows how the architecture maps onto an O-RAN deployment. O-RAN has two RAN Intelligent Controllers (RICs). The Non-Real-Time RIC (Non-RT RIC) hosts applications called rApps, which operate above one second. The Near-Real-Time RIC (Near-RT RIC) hosts xApps, which act between ten milliseconds and one second. The AI functions run as xApps in the Near-RT RIC. The verification agent runs beside them as a dedicated audit xApp and reads their activations. An audit policy rApp in the Non-RT RIC maintains the predicate library and passes it to the audit xApp. It also reports alerts from the audit to the operator and triggers re-certification. Outside O-RAN, the verification agent can run beside a conventional vendor platform, provided that it can read the activations of the audited functions.

Runtime auditing runs in one of two modes. Fig.~\ref{fig:oran}(a) shows monitoring mode. The verification agent observes decisions outside the decision path and records the case assigned to each decision by Step~3. Over a sliding window of recent decisions, it tracks the share of shortcuts and routing failures. It raises an alert when this share exceeds a tolerance set by the operator. A rising share indicates that the reasoning behind a class of decisions has drifted from the baseline, even before the drift causes output errors. Monitoring does not delay the decisions.

Fig.~\ref{fig:oran}(b) shows gating mode. The audit now runs on the decision path. The verification agent blocks each decision classified by Step~3 as a shortcut or a routing failure, and the network falls back to the deterministic protocol-stack logic defined by 3GPP. Gating is therefore safe, because every blocked decision is replaced by standards-compliant behaviour and the network always receives a decision. Gating is also feasible, for two reasons. First, the audited functions run as xApps and act on a timescale of ten milliseconds to one second, which covers RRC decisions and L2 scheduling policies. Second, the check adds little computation. The sparse autoencoder encodes only the baseline features from activations already computed by the forward pass, and the detectors reduce to scalar comparisons.

\subsection{Who Audits the Verification Agent?}
\label{sec:auditor}

A verification agent is useful only if operators can trust its verdicts. Three properties support this trust.

\emph{1) The verdict is based on the 3GPP specifications.} The verification agent matches features against predicates compiled from specification text and removes them alongside random controls of the same size. Both operations follow fixed procedures, and the reference is a set of versioned specifications written by 3GPP working groups. Each verdict can therefore be traced to a specific clause of a specification. The one learned component is the sparse autoencoder, and Section~\ref{sec:challenges} discusses its faithfulness.

\emph{2) The verification agent can be tested against known answers.} We can build two kinds of test agents on purpose. The first is trained to exploit a shortcut. The second is built so that its decisions follow the protocol rule. The correct verdict for each test agent is known in advance, so that the verification agent can be scored on how often it returns that verdict. Such a test is possible for an audit because its correct verdict can be constructed in advance. The correct action in a live network cannot be constructed in this way.

\emph{3) The verification agent fails safely.} It can block a decision, but it cannot make one, and it does not configure the network. A false alarm in gating mode costs the gain expected from the learned function, and a false alarm in monitoring mode costs the time to investigate it. A missed detection leaves the network in the same state as it would be without the audit.

Operators and regulators can further strengthen the audit. They can ratify the predicate library, sign off the certification report before an action agent is admitted, and adjudicate the alerts raised during operation.

\section{Case Study: Auditing an LTE Signalling Model}
\label{sec:casestudy}

\begin{figure*}[!t]
\centering
\includegraphics[width=\textwidth]{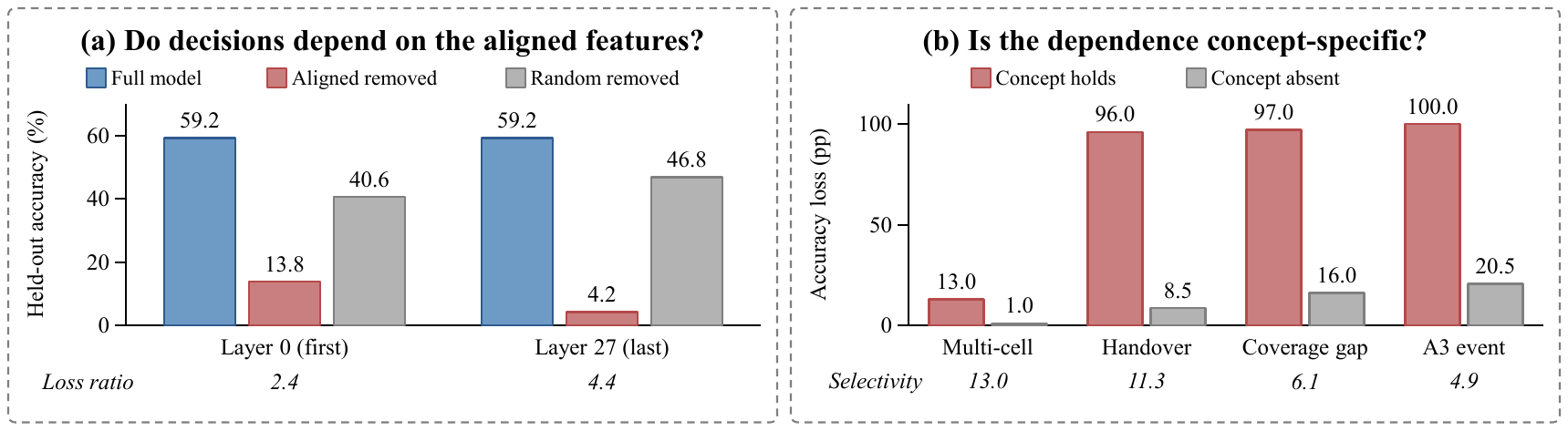}
\caption{Building and testing the audit baseline of an LTE signalling model built on Qwen3-0.6B. (a) Held-out accuracy after removing the aligned features or the same number of random features, at the first and last layers. (b) Accuracy loss on the tasks that each concept governs and on the other tasks after removing that concept's aligned features.}
\label{fig:casestudy}
\end{figure*}

We now apply the principle to a language model that serves as an AI function for LTE signalling. Steps 1 and 2 build the baseline that the audit reads, and the removals of Step~2 also test whether the signal that Step~3 checks carries the meaning that Section~\ref{sec:principle} assigns to it. The model is Qwen3-0.6B adapted with low-rank adaptation (LoRA), and it answers signalling questions of three kinds, namely extracting a value from the measurements, comparing measured quantities, and reasoning over several steps. The tasks are built from radio resource control logs recorded in field tests. The 15 concepts cover five reference signal received power (RSRP) bands over the reporting range of TS~36.133, a neighbour cell stronger than the serving cell, the A3 event of TS~36.331 triggered or not triggered, the handover procedure, three mobility patterns built from these, and three deployment scenarios. Each detector evaluates the RSRP values, event flags and scenario labels recorded with a task. Of the 5,000 tasks, 500 train the adaptation, 3,500 train the sparse autoencoder and 1,000 are held out. The autoencoder is a TopK model with eight times the residual width and 32 active features, trained on the residual stream at eight layers from the first to the last. The model answers 59.2\% of the held-out tasks correctly, against 10.0\% before adaptation, in a single run with a fixed seed.

\textit{1) Locate: which concepts can the audit read?} At the first layer, 151 features of the model, about half of those active, align with at least one concept under the criteria of Step~1. They cover 12 of the 15 concepts, and the three without an aligned feature are the deployment scenarios, which describe the context of a task rather than govern its answer. The proportion of aligned features is not a measure of competence. Across the eight models of the study it stays between 45\% and 52\%, including two models that answer fewer than one task in ten correctly. These 12 concepts set the scope of the audit, because a decision is auditable only when its governing concept has an aligned feature.

\textit{2) Verify: do the decisions depend on the features that the audit reads?} Removing a set of features holds them inactive on every decision, which reproduces the condition that Step~3 flags as a shortcut, namely a decision taken while its governing features are inactive. Removing all the aligned features of the model at the last layer, by projecting their decoder directions out of the residual stream, lowers held-out accuracy from 59.2\% to 4.2\%. Removing the same number of random features, averaged over three draws, lowers it to 46.8\%, as Fig.~\ref{fig:casestudy}(a) shows. The aligned removal therefore costs 55.0 points against 12.4, about four and a half times as much. At the first layer the ratio is about 2.4, and the gap between the two losses therefore widens with depth.

The concept-selective control removes only the features aligned with one concept and compares two groups of 200 tasks, one with the concept present and one with it absent. Because the specification fixes in advance which tasks involve a handover, the two groups are defined independently of the model. Removing the 58 handover features lowers accuracy on the handover tasks from 96\% to 0\% and on the other tasks by 8.5 points, whereas the same number of random features changes accuracy in either group by less than 2 points. The loss on the handover tasks is therefore about 11 times the loss on the other tasks, and we call this ratio the selectivity of the concept. The multi-cell, coverage-gap and A3 concepts give selectivities between about 5 and 13, and the handover concept reaches about 10 on Gemma-2-2B, as Fig.~\ref{fig:casestudy}(b) shows.

These results tie the model's correct decisions to identifiable protocol features. Its correct handover decisions depend on the handover features, and removing those features affects its other decisions far less. The audit relies on this relation at runtime. If the handover features are active and the decision follows them, we can infer that the decision is based on the handover rule. If they are inactive, the decision lacks that basis, and the audit flags it as an unreliable shortcut.

\section{Open Challenges}
\label{sec:challenges}

\textit{Predicate-library coverage.} The library in the case study holds 15 concepts, 12 of them defined over LTE measurements and mobility events, and the audit can check only the decisions governed by these concepts. A clause with a numeric condition on measured values can be compiled directly, as in Fig.~\ref{fig:predlib}. Procedures that depend on protocol state, such as conditional handover and dual connectivity, need a formal model of the procedure, and their detectors need to track the protocol state across messages. Extending the library to NR and to the AI and ML features now entering the standards therefore requires new detectors of both kinds. Automated extraction of predicates from specification text could reduce the effort, provided that each extracted predicate is checked against its clause before it enters the ratified library.

\textit{From recorded signalling to closed-loop control.} The case study evaluates the model on recorded LTE signalling, and it judges each decision on its own. A function that controls a live network differs in three ways. Its inputs arrive as measurement streams, its decisions change the state observed by later decisions, and its errors carry an operational cost. A protocol condition such as a handover trigger develops over successive measurements. An audit of such a function therefore needs to judge each decision together with the decisions before it. The baseline also has to be built from traffic shaped by the function itself, and the scenarios labelled by the detectors then depend on the earlier behaviour of the function. The cost of gating on a real platform has not been measured either, and this cost decides whether gating fits within the latency and energy budget of the platform.

\textit{Keeping the audit reliable.} Three threats can weaken the audit after deployment. The first concerns the sparse autoencoder, the one learned part of the audit. Public benchmarks score such decompositions~\cite{karvonen2025saebench}, but they do not yet establish that the extracted features capture a model's computation~\cite{templeton2026scaling}. The decomposition may also become less faithful once the radio environment moves away from its training conditions. The second threat comes from updates. An update such as fine-tuning may invalidate the baseline, and repeating Steps 1 and 2 after every update is costly. Representational similarity measures such as centred kernel alignment~\cite{kornblith2019similarity} compare two versions of a model, but they are global measures and can stay high while the few baseline features change. A test at the level of these features is therefore needed. The third threat comes from attackers. Runtime auditing flags an action agent manipulated by an attacker, because the manipulation changes its internal features, and this check complements conventional network security~\cite{friha2024llm}. An attacker who knows the baseline is harder to detect. Such an attacker can keep the baseline features active while the agent decides on another basis, and the per-decision check then finds these decisions consistent. Repeating Step~2 on a sample of recent traffic tests whether the decisions still depend on those features, but how far this raises the cost of an attack has not been measured.

\textit{Evaluation and standardisation.} An audit is useful only if its alerts correspond to real faults. Two quantities need to be measured. The first is how often decisions on real traffic are shortcuts or routing failures. The second is how often such a decision is also incorrect, which sets how many alerts an operator receives for each real fault. Shared benchmarks would allow different auditing methods to be compared on the same decisions. They could be built from test agents with known verdicts, as in Section~\ref{sec:auditor}, for each protocol procedure. Operators and regulators also need an agreed format for mechanistic compliance evidence, similar to conformance testing in other areas of engineering. The format would record the version of the predicate library, the baseline of the audited function, the results of the two controls of Step~2, and the tolerance that triggers an alert at runtime. With such a format, audit reports from different vendors could be compared and accepted as evidence of compliance, and engagement with 3GPP and with regulators is essential to agree on it.

\section{Conclusion}
\label{sec:conclusion}

Mobile operators are deploying AI functions to make network decisions. Output testing checks what these functions decide, but a correct output can arise from more than one internal reason. Operators therefore also need evidence about why a function decides as it does. Mechanistic auditing provides this evidence by checking the internal features of a function against machine-verifiable 3GPP specifications. We set out a three-step auditing principle. The first two steps build a baseline of the features that carry each protocol concept, once for each version of a function, and the third step checks every decision against that baseline. We embed the principle in an audit-native architecture. Its verification agent certifies each action agent before deployment and audits the decisions of that agent during operation. A case study on an LTE signalling model shows that removing the features that carry the handover concepts lowers accuracy on handover tasks from 96\% to 0\%. The handover decisions therefore depend on the features checked by the audit. Evaluating the audit on the decisions of agents in a live network is the most direct next step.

\bibliographystyle{IEEEtran}
\bibliography{bare_jrnl}

\vspace{-4pt}
\begin{IEEEbiographynophoto}{Genze Jiang}
is working towards the Ph.D. degree with the Department of Computer Science, Brunel University London, UK.
\end{IEEEbiographynophoto}
\vspace{-15pt}
\begin{IEEEbiographynophoto}{Yizhou Huang}
is working as a research fellow with the Department of Computer Science, Brunel University London, UK.
\end{IEEEbiographynophoto}
\vspace{-15pt}
\begin{IEEEbiographynophoto}{Kezhi Wang}
(Senior Member, IEEE) is a Professor with the Department of Computer Science, Brunel University London, UK. His research interests include wireless communications, mobile edge computing, and machine learning.
\end{IEEEbiographynophoto}

\balance
\vskip 0pt plus 1filll\relax

\end{document}